\documentstyle[art10]{article}
\begin{document}

\begin{titlepage}

\title{Octonion X-product Orbits}

\author{Geoffrey Dixon\thanks{supported by a grant from ..., well, no one,
really.} \\ Department of
Mathematics or Physics \\ Brandeis University \\ Waltham, MA 02254 \\
email: dixon@binah.cc.brandeis.edu
\and Department of Mathematics \\ University of Massachusetts \\ Boston, MA
02125}

\maketitle

\begin{abstract} The octonionic X-product gives to the octonions a flexibility
not found in the other
real division  algebras (reals, complexes, quaternions).  The pattern of this
flexibility is
investigated here.
\end{abstract}

\end{titlepage}

\section*{1. Introduction.}

The inspiration for this article arose from three sources {\bf [1] [2] [3]}.
In {\bf [1]} the
octonionic X-product was introduced, and it was pointed out that although the
7-sphere ($S^{7}$) is
the unique parallelizable manifold not also a group manifold, with the aid of
the X-product $S^{7}$
gains an almost group structure.

In {\bf [2]} the X-product was applied to the 480 renumberings of the
octonionic basis, and it
was pointed out that this set of renumberings actually splits into two sets of
240 renumberings via
an X-product equivalence.  These two sets were dubbed opposites.  In both  {\bf
[1]} and  {\bf [2]}
the ultimate goal was an application of the octonions to string theory, in
which context the
octonions play a natural role.

In {\bf [3]} I presented my own view of division algebra theory and how it
connects to physics.  The
presentation of octonion theory in that monograph is pragmatic, a kind of
get-down-and-get-dirty
mathematics that I find comprehensible and useful.  This article is a result of
the application of
my methods to the octonionic X-product. \\

\section*{2. Four Octonionic Basis Numberings.}

In all that follows the symbols $e_{a}, \; a=1,...,7$, will represent an
orthogonal basis for the
7-dimensional imaginary (pure) octonions, and $e_{0} = 1$ will be the identity.
 There are $7!$
permutations of the indices of the pure octonions, and each gives rise to a
modified copy of {\bf O}
(the real octonion division algebra) with an altered, but still octonionic,
multiplication table.
As it turns out, however, these index rearrangements are not all unique.  In
the end we will find
that there are only 480 distinct multiplication tables for which
$$
e_{a}e_{b}=\pm e_{c}
$$
for all $a,b\in\{0,...,7\}$, and some $a,b$-dependent $c$.

Of all these 480 distinct copies of {\bf O}, there are 4 that are singled out
for their elegance and
symmetry.  These four arise from the following $8 \times 8$ array of binary
numbers (see {\bf [3]}):
\begin{equation}
 O = \left[\begin{array}{cccccccc}
 0 & 0 & 0 & 0 & 0 & 0 & 0 & 0  \\
 0 & 1 & 0 & 0 & 1 & 0 & 1 & 1  \\
 0 & 1 & 1 & 0 & 0 & 1 & 0 & 1  \\
 0 & 1 & 1 & 1 & 0 & 0 & 1 & 0  \\
 0 & 0 & 1 & 1 & 1 & 0 & 0 & 1  \\
 0 & 1 & 0 & 1 & 1 & 1 & 0 & 0  \\
 0 & 0 & 1 & 0 & 1 & 1 & 1 & 0  \\
 0 & 0 & 0 & 1 & 0 & 1 & 1 & 1  \\
\end{array}\right].
\end{equation}
Let $OR_{a}$ be the $a^{th}$ row of $O$, and $OC_{a}$ the $a^{th}$ column.  The
set of rows and
the set of columns are individually closed under binary vector addition
(denoted $\oplus_{2}$).  For
example,
$$
OR_{1}\oplus_{2} OR_{2} = OR_{6};
$$
$$
OC_{1}\oplus_{2} OC_{2} = OC_{4}.
$$
Taking advantage of this closure, I now let the set of rows, or the set of
columns, be bases for
8-dimensional real algebras, and define the following 4 products therefrom
($a\in \{0,1,...,7\}$):
\begin{equation}
\begin{array}{c}
OR_{a}\circ OR_{b} = (-1)^{O_{ab}}(OR_{a}\oplus_{2} OR_{b}); \\
OR_{a}\ast OR_{b} = (-1)^{O_{ba}}(OR_{a}\oplus_{2} OR_{b}); \\
OC_{a}\diamond OC_{b} = (-1)^{O_{ab}}(OC_{a}\oplus_{2} OC_{b}); \\
OC_{a}\star OC_{b} = (-1)^{O_{ba}}(OC_{a}\oplus_{2} OC_{b}). \\
\end{array}
\end{equation}
The power of $(-1)$ out front determines the sign of the result.

Each of the four products in (2) defines an 8-dimensional real algebra from the
array $O$, and each
is isomorphic to the octonion algebra, {\bf O}.  Let $e_{a}$,  be a basis for
{\bf O}, $a\in \{0,1,...,7\}$.  Given any of the products in (2), the $e_{a}$
automatically satisfy
the following useful properties for distinct indices $a,b,c\in \{1,...,7\}$:
\begin{equation}
\fbox{$
\mbox{ if } \; e_{a}e_{b}=\pm e_{c}, \mbox{ then } e_{a+1}e_{b+1}=\pm e_{c+1};
$}
\end{equation}
 and
\begin{equation}
\fbox{$
\mbox{ if } \; e_{a}e_{b}=\pm e_{c}, \mbox{ then } e_{2a}e_{2b}=\pm e_{2c},
$}
\end{equation}
where the indices in (3) and (4) are understood to cycle from 1 to 7 modulo 7.
The multiplication
laws (3) (index cycling) and (4) (index doubling) will only both be valid for
octonion multiplication
rules derived from (2).

Some more general laws, valid for all the 480 renumberings of the $e_{a}$ we
will consider here, are
\begin{equation}
\fbox{$
 e_{a}e_{b}=\pm e_{c} \Longrightarrow e_{c}e_{a}=\pm e_{b}
$}
\end{equation}
(that is, $\{e_{a}, e_{b}, e_{c}\}$ are a quaternionic triple in this case),
and
\begin{equation}
\fbox{$
e_{a}^{2} = -1,
$}
\end{equation}
where again we are assuming $a,b,c\in \{1,...,7\}$ in (5) and (6).

The multiplication laws (3 - 6) are enough to completely determine the octonion
multiplication
tables resulting from the  following four product rules (which arise from the
four
respective products rules defined in (2)):
\begin{equation}
\fbox{$
\begin{array}{cl}
{\bf O}^{+5}: & e_{a}e_{a+1} = e_{a+5}; \\ \\
{\bf O}^{-5}: & e_{a}e_{a+1} = -e_{a+5}; \\ \\
{\bf O}^{+3}: & e_{a}e_{a+1} = e_{a+3}; \\ \\
{\bf O}^{-3}: & e_{a}e_{a+1} = -e_{a+3}. \\
\end{array}$}
\end{equation}
The four copies of {\bf O} that result from the rules (7) and the laws (3 - 6)
are the cornerstones
upon which I shall built the tower of 480 renumberings, and the cement holding
it all together will
be the X-product {\bf [1]}. \\

\section*{3. The 480 Renumberings.}

Given any of the 480 distinct copies of {\bf O}, a
complete multiplication table is determined once one has listed  7 quaternionic
index
triples.  For example, for ${\bf O}^{+5}$, $e_{a}e_{b}=e_{c}$ if $(abc)$ is one
of the following
triples, or any cyclic permutation thereof:
$$
{\bf O}^{+5} \mbox{ triples }: \;
\{(126),(237),(341),(452),(563),(674),(715)\}.
$$

There are simpler, more schematic ways of indicating the same information.  A
common method uses
septagons, but I find the idea of making such a figure with Latex daunting, so
I will use the
following more concise diagrams:
\begin{equation}
\fbox{$
\begin{array}{cl}
{\bf O}^{+5}: & \fbox{$1$} \; 2 \; \fbox{$3$} \; \fbox{$4$} \; 5 \; 6 \; 7; \\
\\
{\bf O}^{-5}: & \fbox{$7$} \; \fbox{$6$} \; 5 \; \fbox{$4$} \; 3 \; 2 \; 1; \\
\\
{\bf O}^{+3}: & \fbox{$1$} \; \fbox{$2$} \; 3 \; \fbox{$4$} \; 5 \; 6 \; 7; \\
\\
{\bf O}^{-3}: & \fbox{$7$} \; 6\; \fbox{$5$} \; \fbox{$4$}\; 3\; 2\; 1. \\
\end{array} $}
\end{equation}
In each case, the quaternionic index triples of the four respective octonionic
algebras are obtained
via a cyclic shifting of the pattern of boxes given for the algebra.  So for
${\bf O}^{+5}$, (134)
is followed by (245), (356), (467), (571), (612), (723), and back to (134).

So, we have 7 index triples, and each triple has 3 cyclic permutations, so
there are $3\times 7 = 21$
pairs of distinct indices $a,b\in \{1,...,7\}$ such that
$$
e_{a}e_{b} = +e_{c}
$$
for some $c\in \{1,...,7\}$.  For each such pair $a$ and $b$ there is a boxed
sequence beginning
with that pair from which the algebra multiplication table is derivable.  For
example, in
${\bf O}^{+5}$, we have $e_{5}e_{7} = e_{1}$, and the sequence
$$
{\bf O}^{+5}: \; \; \fbox{$5$} \; 7 \; \fbox{$2$} \; \fbox{$4$} \; 6 \; 1 \; 3;
\\
$$
results in the same set of index triples for ${\bf O}^{+5}$ as does that in
(8).

Therefore, via this type of rearrangement it is always possible to begin the
boxed sequences of any
of the 480 rearrangements of the indices of {\bf O} with either the pair
(1,2,...) or (2,1,...).
There are
$$
\frac{7!}{21} = 240 = 2(5!)
$$
such reorderings.  And there are {\it two inequivalent} ways of boxing the
triples (modulo cyclic
shifts), those shown it (8).    So there are
$$
2(240) = 480
$$
different multiplication tables resulting from rearrangements of the indices of
the $e_{a}$.   As was
pointed out in {\bf [2]}, these fall into two groups of 240 (different groups
than those above related
by box pattern) related by the octonionic X-product {\bf [1]}. \\ \\

I will consider two other boxed sequences before finishing this section.  In
{\bf [2]} the following
septuplet of index triples is introduced:
\begin{equation}
{\bf O}^{[2]}: \{(123),(145),(167),(264),(257),(347),(356)\}.
\end{equation}
Once you get the hang of it, determining the boxed sequence from this is easy:
\begin{equation}
{\bf O}^{[2]}: \; \; \fbox{$1$}  \; \fbox{$2$} \;6 \; \fbox{$3$} \; 4 \; 5 \;
7. \\
\end{equation}

In {\bf [4]} (which was my introduction to the octonions), a frequently
encountered octonion
multiplication is used with the following septuplet of triples:
\begin{equation}
{\bf O}^{[4]}: \{(123),(174),(275),(376),(165),(246),(354)\}.
\end{equation}
Its boxed sequence is
\begin{equation}
{\bf O}^{[4]}: \; \; \fbox{$1$}  \; 2 \; \fbox{$7$} \; \fbox{$4$} \; 6 \; 3 \;
5. \\
\end{equation} \newpage

\section*{4. The X-Product.}

In what follows I shall use the ${\bf O}^{+5}$ product as a starting point for
all calculations and
X-product variations unless explicitly stated otherwise.

Let $A, B, X \in {\bf O}$, with $X$ a unit octonion ($XX^{\dagger} = 1$) {\bf
[1][3]}.  Define
\begin{equation}
\fbox{$
A\circ_{X}B = (AX)(X^{\dagger}B) = (A(BX))X^{\dagger} = X((X^{\dagger}A)B),
$}
\end{equation}
the X-product of $A$ and $B$.  Because of the nonassociativity of {\bf O},
$A\circ_{X}B \ne AB$ in
general.  But remarkably, for fixed $X$, the algebra ${\bf O}_{X}$ ({\bf O}
endowed with the
X-product) is isomorphic to {\bf O} itself.  Modulo sign change each $X$ gives
rise to a distinct
copy of {\bf O}, so the orbit of copies of {\bf O} arising from any given
starting copy is
\begin{equation}
S^{7}/Z_{2} = RP^{7},
\end{equation}
the manifold obtained from the 7-sphere by identifying opposite points.
Moreover, composition of
X-products is yet another X-product.  That is, if
$$
X,Y\in S^{7}\subset {\bf O},
$$
then
\begin{equation}
\begin{array}{rcl}
AB \stackrel{X}{\longrightarrow}A\circ_{X}B &=& (AX)(X^{\dagger}B)
\stackrel{Y}{\longrightarrow} \\
(A\circ_{X}Y)\circ_{X}(Y^{\dagger}\circ_{X}B) &=&
  [((AX)(X^{\dagger}Y))X][X^{\dagger}((Y^{\dagger}X)(X^{\dagger}B))] \\
&=& [((A(YX))X^{\dagger})X][X^{\dagger}(X((X^{\dagger}Y^{\dagger})B))] \\
&=& [A(YX)][(X^{\dagger}Y^{\dagger})B] \\
&=& A\circ_{(YX)}B, \\
\end{array}
\end{equation}
using the fact that for $X\in S^{7}$,
$$
(UX^{\dagger})X = U = X^{\dagger}(XU)
$$
for all $U\in {\bf O}$.  (This would seem to endow $RP^{7}$ with a Lie group
structure, but in
composing with yet a third element of $S^{7}$ one runs into the
nonassociativity of {\bf O}, which
spoils the game.) \\ \\

Clearly in general the result of the X-product $e_{a}\circ_{X} e_{b}, \; 0\ne
a\ne b\ne 0,$ will be
a linear combination of $e_{c}, \; c\in \{1,...,7\}$ (it is not difficult to
prove that such a
product can not have any terms linear in the identity).  There are some $X$,
however, such that for
all $a,b\in \{1,...,7\}$, there will be a particular $c\in \{1,...,7\}$
satisfying
$$
e_{a}\circ_{X} e_{b} = e_{c}.
$$
Starting from ${\bf O}^{+5}$, any such $X$ resides in one of the following
sets:
\begin{equation}
\fbox{$
\begin{array}{cl}
\Xi^{+5}_{0} = & \{\pm e_{a}\}, \\ \\
\Xi^{+5}_{1} = & \{(\pm e_{a}\pm e_{b})/\sqrt{2}: a,b \mbox{ distinct}\}, \\ \\
\Xi^{+5}_{2} = & \{(\pm e_{a}\pm e_{b}\pm e_{c}\pm e_{d})/2: a,b,c,d \mbox{
distinct}, \;
e_{a}(e_{b}(e_{c}e_{d}))=\pm 1\}, \\ \\
\Xi^{+5}_{3} = & \{(\sum_{a=0}^{7}\pm e_{a})/\sqrt{8}:
\mbox{ odd number of +'s}  \}, \\ \\
& a,b,c,d\in\{0,...,7\}, e_{a},e_{b},e_{c},e_{d}\in {\bf O}^{+5}. \\
\end{array}
$}
\end{equation} \\

\begin{itemize}
\item{NOTE:} These sets not general.  They will work for ${\bf O}^{\pm 5}$, and
certain of their
X-product variants.  See section 5.  (Quite frankly, I don't know how many such
sets there are for
the total of 480 {\bf O}'s, but not 480.) \\
\end{itemize}

In {\bf [3]} I explicitly calculated the effect of the X-product (13).  For
example, for a general
$$
X =
X^{0}+X^{1}e_{1}+X^{2}e_{2}+X^{3}e_{3}+X^{4}e_{4}+X^{5}e_{5}+X^{6}e_{6}+X^{7}e_{7}\in S^{7},
$$
\begin{equation}
\fbox{$
\begin{array}{l} e_{1}\circ_{X} e_{2} = \\ \\
((X^{0})^{2}+(X^{1})^{2}+(X^{2})^{2}+(X^{6})^{2}
-(X^{3})^{2}-(X^{4})^{2}-(X^{5})^{2}-(X^{7})^{2})e_{6} \\ \\
  +2(X^{0}X^{5}+X^{1}X^{7}-X^{2}X^{4}+X^{3}X^{6})e_{3} \\ \\
  +2(-X^{0}X^{7}+X^{1}X^{5}+X^{2}X^{3}+X^{4}X^{6})e_{4} \\ \\
  +2(-X^{0}X^{3}-X^{1}X^{4}-X^{2}X^{7}+X^{5}X^{6})e_{5} \\ \\
  +2(X^{0}X^{4}-X^{1}X^{3}+X^{2}X^{5}+X^{7}X^{6})e_{7}. \\
\end{array}
$}
\end{equation}
Because (17) arises from ${\bf O}^{+5}$, all the other possible $e_{a}\circ_{X}
e_{b}$ are derivable
from (17) via index cycling and doubling (in both cases, the index $0$ is left
out, and only the
indices $a=1,...,7$ are subject to the cycling and doubling operations).  Some
other products that
will prove useful later are: \newpage
$$
\begin{array}{c}
\mbox{\Large Table of X-Products for ${\bf O}^{+5}$} \\ \\
\fbox{$
\begin{array}{cl} e_{3}\circ_{X} e_{4} = &
((X^{0})^{2}+(X^{3})^{2}+(X^{4})^{2}+(X^{1})^{2}
-(X^{2})^{2}-(X^{5})^{2}-(X^{6})^{2}-(X^{7})^{2})e_{1} \\ \\
 &   +2(X^{0}X^{7}+X^{3}X^{2}-X^{4}X^{6}+X^{5}X^{1})e_{5} \\ \\
 &   +2(-X^{0}X^{2}+X^{3}X^{7}+X^{4}X^{5}+X^{6}X^{1})e_{6} \\ \\
 &   +2(-X^{0}X^{5}-X^{3}X^{6}-X^{4}X^{2}+X^{7}X^{1})e_{7} \\ \\
 &   +2(X^{0}X^{6}-X^{3}X^{5}+X^{4}X^{7}+X^{2}X^{1})e_{2}. \\
\end{array}
$} \\
\fbox{$
\begin{array}{cl} e_{4}\circ_{X} e_{5} = &
((X^{0})^{2}+(X^{4})^{2}+(X^{5})^{2}+(X^{2})^{2}
-(X^{1})^{2}-(X^{3})^{2}-(X^{6})^{2}-(X^{7})^{2})e_{2} \\ \\
&  +2(X^{0}X^{1}+X^{4}X^{3}-X^{5}X^{7}+X^{6}X^{2})e_{6} \\ \\
&  +2(-X^{0}X^{3}+X^{4}X^{1}+X^{5}X^{6}+X^{7}X^{2})e_{7} \\ \\
&  +2(-X^{0}X^{6}-X^{4}X^{7}-X^{5}X^{3}+X^{1}X^{2})e_{1} \\ \\
&  +2(X^{0}X^{7}-X^{4}X^{6}+X^{5}X^{1}+X^{3}X^{2})e_{3}. \\
\end{array}
$} \\
\fbox{$
\begin{array}{cl} e_{5}\circ_{X} e_{6} = &
((X^{0})^{2}+(X^{5})^{2}+(X^{6})^{2}+(X^{3})^{2}
-(X^{2})^{2}-(X^{4})^{2}-(X^{7})^{2}-(X^{1})^{2})e_{3} \\ \\ &
+2(X^{0}X^{2}+X^{5}X^{4}-X^{6}X^{1}+X^{7}X^{3})e_{7} \\ \\ &
+2(-X^{0}X^{4}+X^{5}X^{2}+X^{6}X^{7}+X^{1}X^{3})e_{1} \\ \\ &
+2(-X^{0}X^{7}-X^{5}X^{1}-X^{6}X^{4}+X^{2}X^{3})e_{2} \\ \\ &
+2(X^{0}X^{1}-X^{5}X^{7}+X^{6}X^{2}+X^{4}X^{3})e_{4}. \\
\end{array}
$} \\
\fbox{$
\begin{array}{cl} e_{6}\circ_{X} e_{7} = &
((X^{0})^{2}+(X^{6})^{2}+(X^{7})^{2}+(X^{4})^{2}
-(X^{3})^{2}-(X^{5})^{2}-(X^{1})^{2}-(X^{2})^{2})e_{4} \\ \\ &
+2(X^{0}X^{3}+X^{6}X^{5}-X^{7}X^{2}+X^{1}X^{4})e_{1} \\ \\ &
+2(-X^{0}X^{5}+X^{6}X^{3}+X^{7}X^{1}+X^{2}X^{4})e_{2} \\ \\ &
+2(-X^{0}X^{1}-X^{6}X^{2}-X^{7}X^{5}+X^{3}X^{4})e_{3} \\ \\ &
+2(X^{0}X^{2}-X^{6}X^{1}+X^{7}X^{3}+X^{5}X^{4})e_{5}. \\
\end{array}
$} \\
\end{array}
$$
$$
\begin{array}{c}
\mbox{\Large ${\bf O}^{+5}$ X-Products continued} \\ \\
\fbox{$
\begin{array}{cl} e_{1}\circ_{X} e_{3} = &
((X^{0})^{2}+(X^{1})^{2}+(X^{3})^{2}+(X^{4})^{2}
-(X^{2})^{2}-(X^{5})^{2}-(X^{6})^{2}-(X^{7})^{2})e_{4} \\ \\
 &   +2(X^{0}X^{2}+X^{1}X^{6}-X^{3}X^{7}+X^{5}X^{4})e_{5} \\ \\
 &   +2(-X^{0}X^{6}+X^{1}X^{2}+X^{3}X^{5}+X^{7}X^{4})e_{7} \\ \\
 &   +2(-X^{0}X^{5}-X^{1}X^{7}-X^{3}X^{6}+X^{2}X^{4})e_{2} \\ \\
 &   +2(X^{0}X^{7}-X^{1}X^{5}+X^{3}X^{2}+X^{6}X^{4})e_{6}. \\
\end{array}
$} \\
\fbox{$
\begin{array}{cl} e_{2}\circ_{X} e_{4} = &
((X^{0})^{2}+(X^{2})^{2}+(X^{4})^{2}+(X^{5})^{2}
-(X^{1})^{2}-(X^{3})^{2}-(X^{6})^{2}-(X^{7})^{2})e_{5} \\ \\
 &   +2(X^{0}X^{3}+X^{2}X^{7}-X^{4}X^{1}+X^{6}X^{5})e_{6} \\ \\
 &   +2(-X^{0}X^{7}+X^{2}X^{3}+X^{4}X^{6}+X^{1}X^{5})e_{1} \\ \\
 &   +2(-X^{0}X^{6}-X^{2}X^{1}-X^{4}X^{7}+X^{3}X^{5})e_{3} \\ \\
 &   +2(X^{0}X^{1}-X^{2}X^{6}+X^{4}X^{3}+X^{7}X^{5})e_{7}. \\
\end{array}
$} \\
\fbox{$
\begin{array}{cl} e_{3}\circ_{X} e_{5} = &
((X^{0})^{2}+(X^{3})^{2}+(X^{5})^{2}+(X^{6})^{2}
-(X^{1})^{2}-(X^{2})^{2}-(X^{4})^{2}-(X^{7})^{2})e_{6} \\ \\
 &   +2(X^{0}X^{4}+X^{3}X^{1}-X^{5}X^{2}+X^{7}X^{6})e_{7} \\ \\
 &   +2(-X^{0}X^{1}+X^{3}X^{4}+X^{5}X^{7}+X^{2}X^{6})e_{2} \\ \\
 &   +2(-X^{0}X^{7}-X^{3}X^{2}-X^{5}X^{1}+X^{4}X^{6})e_{4} \\ \\
 &   +2(X^{0}X^{2}-X^{3}X^{7}+X^{5}X^{4}+X^{1}X^{6})e_{1}. \\
\end{array}
$} \\
\fbox{$
\begin{array}{cl} e_{4}\circ_{X} e_{6} = &
((X^{0})^{2}+(X^{4})^{2}+(X^{6})^{2}+(X^{7})^{2}
-(X^{1})^{2}-(X^{2})^{2}-(X^{3})^{2}-(X^{5})^{2})e_{7} \\ \\
 &   +2(X^{0}X^{5}+X^{4}X^{2}-X^{6}X^{3}+X^{1}X^{7})e_{1} \\ \\
 &   +2(-X^{0}X^{2}+X^{4}X^{5}+X^{6}X^{1}+X^{3}X^{7})e_{3} \\ \\
 &   +2(-X^{0}X^{1}-X^{4}X^{3}-X^{6}X^{2}+X^{5}X^{7})e_{5} \\ \\
 &   +2(X^{0}X^{3}-X^{4}X^{1}+X^{6}X^{5}+X^{2}X^{7})e_{2}. \\
\end{array}
$} \\
\end{array}
$$ \newpage

Clearly this is not a complete set, but it will suffice to construct a few
examples. \\

\begin{itemize}
\item{EXAMPLE 1: }  $X = (e_{0}- e_{1}- e_{2}- e_{3}- e_{4}- e_{5}- e_{6}-
e_{7})/\sqrt{8}$. \\
Note that because this $X$ is invariant under both index cycling and index
doubling, the product
$e_{a}\circ_{X}e_{b}$ will have the index cycling and doubling properties
shared by the products
(7).  Using the tables given above we can see that the index triples associated
with this
modification of the ${\bf O}^{+5}$ product are those of ${\bf O}^{+3}$.  That
is,
\begin{equation}
\fbox{$
\mbox{ if } X = (e_{0}- e_{1}- e_{2}- e_{3}- e_{4}- e_{5}- e_{6}-
e_{7})/\sqrt{8},
\mbox{ then } {\bf O}_{X}^{+5} = {\bf O}^{+3}.
$}
\end{equation}
Since $XX^{\dagger}=1$,
\begin{equation}
\fbox{$
\mbox{ if } X = (e_{0}- e_{1}- e_{2}- e_{3}- e_{4}- e_{5}- e_{6}-
e_{7})/\sqrt{8},
\mbox{ then } {\bf O}_{X^{\dagger}}^{+3} = {\bf O}^{+5}.
$}
\end{equation}
Therefore ${\bf O}^{+5}$ and ${\bf O}^{+3}$ are part of the same orbit of
octonion X-product
variants. \\

\item{EXAMPLE 2: } $X = ( e_{1}- e_{2}- e_{4}- e_{7})/2$. \\
In this case,
$$
X^{0} =X^{3} =X^{5} =X^{6} = 0; \; \; X^{1} =-X^{2} =-X^{4} =-X^{7} =
\frac{1}{2}.
$$
Plug these values into the general form for $e_{1}\circ_{X}e_{2}$ in (17) and
get
$$
e_{1}\circ_{X}e_{2} = -e_{3}.
$$
Therefore (132) is a quaternionic index triple for this ${\bf O}_{X}$.
Likewise, from the general
$e_{4}\circ_{X}e_{5}$ table,
$$
e_{4}\circ_{X}e_{5} = -e_{1},
$$
so (154) is another such triple.  Carrying on in this way leads to a complete
set, with corresponding
boxed sequence:
\begin{equation}
\begin{array}{c}
(321),(541),(761),(462),(752),(743),(653). \\ \\
 \fbox{$2$}  \; 1 \; \fbox{$7$} \; \fbox{$5$} \; 4 \; 3 \; 6. \\
\end{array}
\end{equation}
Compare this set of triples to the set (9) for ${\bf O}^{[2]}$.  Each of the
triples in (20) is
reversed with respect to a corresponding triple in (9).  Following {\bf [2]} I
shall call the copy
of {\bf O} generated from (20) the opposite of that generated from (9) (${\bf
O}^{[2]}$).  Let's
denote it $\underline{{\bf O}}^{[2]}$, the underline signifying opposite.
\end{itemize}

So
$$
\underline{{\bf O}}^{[2]} = {\bf O}_{X}^{+5}.
$$
That is, they are a part of the same orbit.  What about
${\bf O}^{[2]}$ itself?  It turns out that ${\bf O}^{+5}$ and ${\bf O}^{[2]}$
are not in the same
orbit.  Consider ${\bf O}^{+5}$ and ${\bf O}^{-5}$ as an example.  The
quaternionic triples of
${\bf O}^{-5}$ are the reverse of those for ${\bf O}^{+5}$.  (So
$$
\underline{{\bf O}}^{+5} = {\bf O}^{-5}.)
$$
Therefore,
$$
e_{1}e_{2} = e_{6} \mbox{ in } {\bf O}^{+5} \mbox{ implies }
e_{1}e_{2} = -e_{6} \mbox{ in } {\bf O}^{-5}.
$$
For this to happen via an X-product, we can see from (17) that
$$
\begin{array}{r}
(X^{0})^{2}+(X^{1})^{2}+(X^{2})^{2}+(X^{6})^{2}
-(X^{3})^{2}-(X^{4})^{2}-(X^{5})^{2}-(X^{7})^{2} = -1, \\ \\
  X^{0}X^{5}+X^{1}X^{7}-X^{2}X^{4}+X^{3}X^{6} =0, \\ \\
  -X^{0}X^{7}+X^{1}X^{5}+X^{2}X^{3}+X^{4}X^{6}  =0, \\ \\
  -X^{0}X^{3}-X^{1}X^{4}-X^{2}X^{7}+X^{5}X^{6} =0,  \\ \\
  X^{0}X^{4}-X^{1}X^{3}+X^{2}X^{5}+X^{7}X^{6} =0. \\
\end{array}
$$
These equations can be satisfied in several ways.  For example, set $X^{7}=1$,
and
$X^{a}=0, \; a=0,...,6$.  But whatever value of $X$ we choose must reverse not
only this product,
but every other ${\bf O}^{+5}$ product.  In particular, this implies that each
of the following
eight equations must be satisfied:
$$
\left[
\begin{array}{rrrrrrrr}
1 & 1 & 1 & 1 & 1 & 1 & 1 & 1 \\
1 & -1 & -1 & -1 & 1 & -1 & 1 & 1 \\
1 & 1 & -1 & -1 & -1 & 1 & -1 & 1 \\
1 & 1 & 1 & -1 & -1 & -1 & 1 & -1 \\
1 & -1 & 1 & 1 & -1 & -1 & -1 & 1 \\
1 & 1 & -1 & 1 & 1 & -1 & -1 & -1 \\
1 & -1 & 1 & -1 & 1 & 1 & -1 & -1 \\
1 & -1 & -1 & 1 & -1 & 1 & 1 & -1 \\
\end{array}\right]
\left[
\begin{array}{r}
(X^{0})^{2}  \\ (X^{1})^{2} \\ (X^{2})^{2} \\ (X^{6})^{2} \\ (X^{3})^{2}  \\
(X^{4})^{2} \\
(X^{5})^{2} \\ (X^{7})^{2} \\
\end{array}\right] =
\left[
\begin{array}{r}
1 \\ -1 \\ -1 \\ -1 \\ -1 \\ -1 \\ -1 \\ -1 \\
\end{array}\right].
$$
The inverse of that square matrix is $\frac{1}{8}$ times its transpose.
Therefore a solution is
easily obtained, and in particular it implies that
$$
(X_{0})^{2} = -\frac{3}{4},
$$
which of course is not possible for the real algebra {\bf O}. (As it turns out,
even
the complexification of {\bf O} wouldn't help in the end.)

\begin{itemize}
\item{EXAMPLE 3: } $X = (e_{0}- e_{1}- e_{2}- e_{3}- e_{4}- e_{5}+ e_{6}+
e_{7})/\sqrt{8}$. \\
In this case, using (17) and the other ${\bf O}^{+5}$ X-product tables, we
find:
$$
e_{1}\circ_{X}e_{2}=-e_{3}, \; e_{3}\circ_{X}e_{4}=e_{5}, \;
e_{5}\circ_{X}e_{6}=e_{1}, \;
e_{6}\circ_{X}e_{7}=e_{3}, \; e_{2}\circ_{X}e_{4}=-e_{6}.
$$
By cyclically shifting the $e_{4}\circ_{X}e_{6}$ table up by $1$, and the
$e_{1}\circ_{X}e_{2}$
table down by $1$, we get
$$
e_{5}\circ_{X}e_{7}=e_{2}, \; \; e_{7}\circ_{X}e_{1}=e_{4}.
$$
Therefore, a complete set of quaternionic triples for this case, with
corresponding boxed sequence,
is:
\begin{equation}
\begin{array}{c}
(321),(471),(572),(673),(561),(642),(453), \\ \\
\fbox{$2$}  \;  \fbox{$1$} \; 5\; \fbox{$3$} \; 6 \; 4 \; 7. \\
\end{array}
\end{equation}
These are all opposite those for ${\bf O}^{[4]}$ listed in (11).  Therefore, in
this case,
$$
{\bf O}^{+5}_{X} = \underline{{\bf O}}^{[4]}.
$$ \\

\item{NOTE: } In general, if {\bf O} and ${\bf O}'$ are in the same X-product
orbit, then {\bf
O} and
$\underline{\bf O}'$ are not.  There are two orbits all together, one arising
from ${\bf
O}^{+5}$ and containing ${\bf O}^{+3}$ (${\bf O}rbit^{+}$), and one arising
from ${\bf O}^{-5}$ and
containing
${\bf O}^{-3}$ (${\bf O}rbit^{-}$).
\end{itemize}

\section*{5. Two X-Product Orbits.}

Look again at the sets $\Xi^{+5}_{a}, a=0,1,2,3$, elements of which cause index
rearrangements of
${\bf O}^{\pm 5}$ via X-product variation.  Modulo
sign change,
\begin{equation}
\fbox{$
\begin{array}{l}
\mbox{ order }\Xi^{+5}_{0} = 8, \\ \\
\mbox{ order }\Xi^{+5}_{1} = 56, \\ \\
\mbox{ order }\Xi^{+5}_{2} = 112, \\ \\
\mbox{ order }\Xi^{+5}_{3} = 64. \\
\end{array}
$}
\end{equation}
So there are 240 rearrangements arising from these elements all together.
Hence each of the two
X-product orbits,
${\bf O}rbit^{\pm}$, contains 240 of the 480 octonion index rearrangements.  In
each orbit there are
120 rearrangements using the ${\bf O}^{+5}$ (and ${\bf O}^{-3}$)  pattern of
boxes (denote these
${\bf O}rbit_{0}^{\pm}$, a discrete subset of ${\bf O}rbit^{\pm}$),
\begin{equation}
\fbox{}\_\fbox{}\fbox{}\_\_\_ \; ,
\end{equation}
 and
120 using the ${\bf O}^{-5}$ (and ${\bf O}^{+3}$)  pattern of boxes (denote
these
${\bf O}rbit_{1}^{\pm}$),
\begin{equation}
\fbox{}\fbox{}\_\fbox{}\_\_\_ \; .
\end{equation}

Starting from ${\bf O}^{+5}$, the 120 distinct X-products arising from elements
of
$\Xi^{+5}_{0}\cup\Xi^{+5}_{2}$ will result in  variants ${\bf O}^{+5}_{X}$
sharing ${\bf O}^{+5}$'s
box pattern (23).  For example, the boxed sequence of ${\bf O}^{+5}_{X}$,
$X=(e_{1}-e_{2}-e_{4}-e_{7})/2
\in
\Xi^{+5}_{2}$, given in (2), has the same box pattern as ${\bf O}^{+5}$ itself.
  (Remember that for
any given reordering of the indices only one of the box patterns is possible.)

Starting from  ${\bf O}^{+5}$, the 120  distinct X-products arising from
elements of
$\Xi^{+5}_{1}\cup\Xi^{+5}_{3}$ will result in variants ${\bf O}^{+5}_{X}$
having the  box pattern
(24).  For example, the boxed sequence of ${\bf O}^{+5}_{X} = {\bf O}^{+3}$,
$X=(1-e_{1}-e_{2}-e_{3}-e_{4}-e_{5}-e_{6}-e_{7})/\sqrt{8} \in
\Xi^{+5}_{3}$, has pattern (24).

Schematically,
\begin{equation}
\begin{array}{ccccc} {\bf O}rbit_{1}^{+}  &
\longleftarrow\Xi^{+5}_{1}\cup\Xi^{+5}_{3} \mbox{---} &
{\bf O}^{+5} &
\mbox{---}\Xi^{+5}_{0}\cup\Xi^{+5}_{2} \longrightarrow & {\bf O}rbit_{0}^{+} \\
\\
\uparrow & & \uparrow & & \uparrow \\
\mbox{opposite} && \mbox{opposite} && \mbox{opposite} \\
\downarrow &  & \downarrow & & \downarrow \\ \\ {\bf O}rbit_{0}^{-} &
\longleftarrow\Xi^{+5}_{1}\cup\Xi^{+5}_{3}
\mbox{---} & {\bf O}^{-5} &
\mbox{---}\Xi^{+5}_{0}\cup\Xi^{+5}_{2} \longrightarrow  & {\bf O}rbit_{1}^{-}
\\
\end{array}
\end{equation}

\begin{itemize}
\item{NOTE: }  If $X \not\in
\Xi^{+5}_{0}\cup\Xi^{+5}_{1}\cup\Xi^{+5}_{2}\cup\Xi^{+5}_{3}$, then
${\bf O}^{+5}_{X}$ is not any of the simple index rearrangements of ${\bf
O}^{+5}$, which are after
all just discrete points in the full $RP^{7}$ orbit of ${\bf O}^{+5}$. \\
\end{itemize}

The  set $\Xi^{+5}_{0}\cup\Xi^{+5}_{1}\cup\Xi^{+5}_{2}\cup\Xi^{+5}_{3}$ is
${\bf O}^{\pm 5}$
specific.  Starting from a general ${\bf O}^{+5}_{X}$, for example, we know
that
$$
({\bf O}^{+5}_{X})_{X^{\dagger}} = {\bf O}_{(X^{\dagger}X)}^{+5} = {\bf
O}^{+5},
$$
and
$$
(({\bf O}^{+5}_{X})_{X^{\dagger}})_{Y} = {\bf O}^{+5}_{Y} = ({\bf
O}^{+5}_{X})_{(YX^{\dagger})}
$$
(see (15)).  Therefore, for $({\bf O}^{+5}_{X})_{Z}$ to be an index
rearrangement of
${\bf O}^{+5}_{X}$, $Z$ must satisfy
\begin{equation}
\fbox{$
Z \in
(\Xi^{+5}_{0}\cup\Xi^{+5}_{1}\cup\Xi^{+5}_{2}\cup\Xi^{+5}_{3})X^{\dagger}.
$}
\end{equation}

Therefore, for example, had we started with ${\bf O}^{+3} = {\bf O}^{+5}_{X}$,
where
$X = (1-e_{1} -...-e_{7})/\sqrt{8}$, so that
$$
X^{\dagger} = (1+e_{1} +...+e_{7})/\sqrt{8},
$$
then the set
$(\Xi^{+5}_{0}\cup\Xi^{+5}_{1}\cup\Xi^{+5}_{2}\cup\Xi^{+5}_{3})X^{\dagger}$
can be
broken up into subsets
\begin{equation}
\fbox{$
\begin{array}{cl}
\Xi^{+3}_{0} = & \{\pm e_{a}\}, \\ \\
\Xi^{+3}_{1} = & \{(\pm e_{a}\pm e_{b})/\sqrt{2}: a,b \mbox{ distinct}\}, \\ \\
\Xi^{+3}_{2} = & \{(\pm e_{a}\pm e_{b}\pm e_{c}\pm e_{d})/2: a,b,c,d \mbox{
distinct}, \;
e_{a}(e_{b}(e_{c}e_{d}))=\pm 1\}, \\ \\
\Xi^{+3}_{3} = & \{(\sum_{a=0}^{7}\pm e_{a})/\sqrt{8}:
\mbox{ even number of +'s}  \}, \\ \\  & a,b,c,d\in\{0,...,7\},
e_{a},e_{b},e_{c},e_{d}\in {\bf
O}^{+3}. \\
\end{array}
$}
\end{equation} \\
Everything that was said for ${\bf O}^{+5}$ and the $\Xi^{+5}_{m}$ holds true
in an analogous way
for ${\bf O}^{+3}$ and the $\Xi^{+3}_{m}$.

\section*{6. Conclusion.}

In the case of the quaternions {\bf Q} there are also "opposites", but because
{\bf Q} is
associative there are no X-product variations.  The quaternion basis with
multiplication table
determined by
$$
q_{1}q_{2} = q_{3}
$$
has an opposite representation with a multiplication table determined by
$$
q_{2}q_{1} = -q_{1}q_{2} = q_{3}.
$$
So we drop from 480 variations down to 2.

There is obviously no way to vary the complex numbers {\bf C}, the smallest of
the three hypercomplex
real division algebras.

Finally, in {\bf [1][2][3]} interest in the division algebras arose from their
evidently intimate
connection to our physical reality.  In a future article I will investigate the
potential and
consequences of gauging the X-product.

\end{document}